\documentclass[aps,physrev,reprint,superscriptaddress,longbibliography]{revtex4-2}

\usepackage{amsmath}
\usepackage{graphicx}
\usepackage[separate-uncertainty]{siunitx}
\usepackage{mhchem}

\graphicspath{{figures/}}

\renewcommand{\vec}[1]{\mathbf{#1}}
\newcommand{\abs}[1]{\left\vert #1 \right\vert}
\newcommand{\real}[1]{\Re\!\left\{ #1 \right\}}
\DeclareSIUnit\pixel{px}

\begin{document}

\title{Ultrasonic chaining of emulsion droplets}

\author{Mohammed A. Abdelaziz} \affiliation{Department of Physics and
  Center for Soft Matter Research, New York University, New York, NY
  10003, USA}

\author{Jairo A. Diaz} \altaffiliation{Current address: Department of
  Chemical Engineering, Rochester Institute of Technology, Rochester,
  NY 14623, USA} \affiliation{Department of Physics and Center for
  Soft Matter Research, New York University, New York, NY 10003, USA}
\affiliation{Department of Chemical and Biomolecular Engineering, New
  York University, New York, NY 11201, USA}

\author{Jean-Luc Aider} \affiliation{Laboratoire PMMH (Physique et
  M\'ecanique des Milieux H\'et\'erog\`enes), UMR7636 CNRS, ESPCI
  Paris, Sorbonne Universit\'es, Paris Sciences Lettres, 1 rue
  Jussieu, 75005 Paris, France}

\author{David J. Pine} \affiliation{Department of Physics and Center
  for Soft Matter Research, New York University, New York, NY 10003,
  USA} \affiliation{Department of Chemical and Biomolecular
  Engineering, New York University, New York, NY 11201, USA}

\author{David G. Grier} \affiliation{Department of Physics and Center
  for Soft Matter Research, New York University, New York, NY 10003,
  USA}

\author{Mauricio Hoyos} \affiliation{Laboratoire PMMH (Physique et
  M\'ecanique des Milieux H\'et\'erog\`enes), UMR7636 CNRS, ESPCI
  Paris, Sorbonne Universit\'es, Paris Sciences Lettres, 1 rue
  Jussieu, 75005 Paris, France}

\date{\today}

\begin{abstract}
  Emulsion droplets trapped in an ultrasonic levitator behave in two
  ways that solid spheres do not: (1) Individual droplets spin rapidly
  about an axis parallel to the trapping plane, and (2) coaxially
  spinning droplets form long chains aligned with their common axis of
  rotation.  Acoustically-organized chains interact hydrodynamically,
  either to merge into longer chains or to form three-dimensional
  bundles of chains.  Solid spheres, by contrast, form close-packed
  planar crystals drawn together by the sound-mediated secondary
  Bjerknes interaction.  We demonstrate the chain-to-crystal
  transition with a model system in which fluid emulsion droplets can
  be photopolymerized into solid spheres without significantly
  changing other material properties.  The behavior of this
  experimental system is quantitatively consistent with an
  acoustohydrodynamic model for spinning spheres in an acoustic
  levitator.  This study therefore introduces acoustically-driven
  spinning as a mechanism for guiding self-organization of
  acoustically levitated matter.
\end{abstract}

\maketitle

\section{Introduction}
\label{sec:introduction}

\begin{figure*}
  \centering \includegraphics[width=0.9\textwidth]{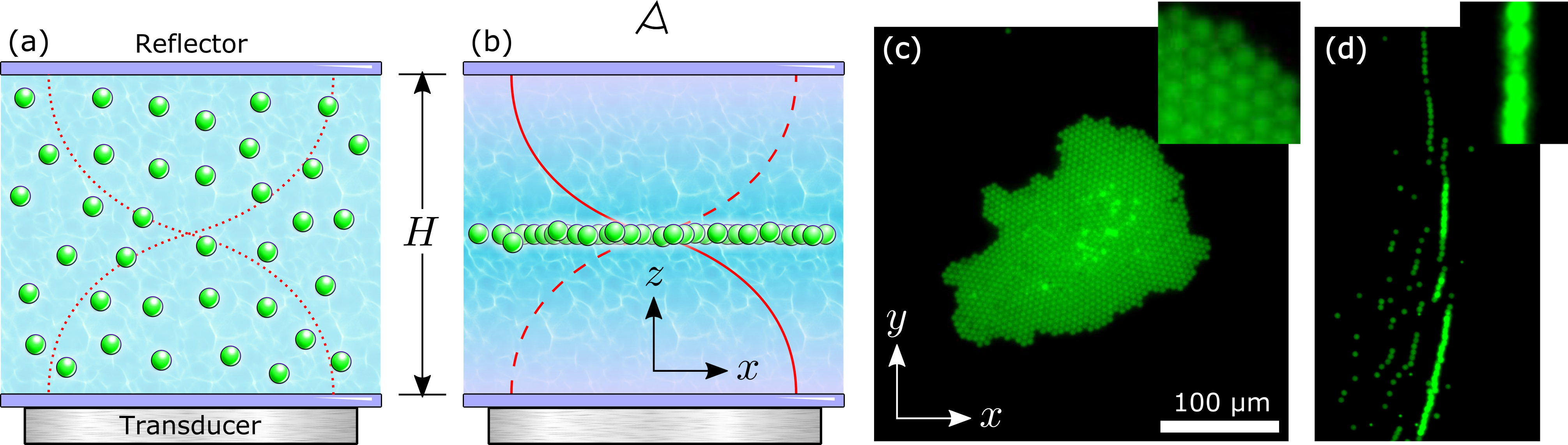}
  \caption{(a) Colloidal particles are dispersed in a horizontal layer
    of water confined between an ultrasonic transducer and a parallel
    transparent reflector separated by half of a wave length,
    $H = \lambda/2$.  (b) The node in the acoustic pressure field
    along the mid-plane acts as an acoustic trap for the dispersed
    particles. (c) Fluorescence microscopy image of solid TPM spheres
    trapped at the levitator's mid-plane and assembled into a
    two-dimensional crystal by secondary Bjerknes forces. Inset is
    $4\times$ magnified.  (d) Fluorescence microscopy image of
    monodisperse droplets of TPM oil levitating at the midplane and
    forming rapidly rotating chains. The individual spheres' axis of
    rotation is aligned with the axis of the chain.}
  \label{fig:phenomenology}
\end{figure*}

Acoustic standing waves exert forces that can be strong enough to
levitate objects against gravity \cite{brandt2001suspended} as
illustrated schematically in Fig.~\ref{fig:phenomenology}(a) and (b).
In addition to experiencing this primary acoustic radiation force
(ARF), objects trapped in a standing wave also interact with each
other through the sound waves that they scatter.  Scattered waves
interfere with the standing wave and with each other to mediate
secondary Bjerknes forces
\cite{silva2014acoustic,garcia2014experimental,castro2016determination}
that generally are attractive at separations smaller than the
wavelength of sound.  Levitated ensembles of objects therefore tend to
organize themselves into planar clusters
\cite{rabaud2011acoustically}.  When applied to monodisperse spheres,
this mechanism creates close-packed crystals
\cite{rabaud2011acoustically,caleap2014acoustically}, such as the
example in Fig.~\ref{fig:phenomenology}(c).

Here, we report a distinct mode of sound-mediated organization that
arises when acoustic levitation is applied to an emulsion of
monodisperse oil droplets in water.  Rather than forming crystals,
these droplets align themselves into long single-file chains, such as
the examples in Fig.~\ref{fig:phenomenology}(d).  Video microscopy of
the chains suggests that the individual droplets rotate rapidly with a
common axis of rotation that is aligned along the chain.

Polymerizing the droplets into solid spheres essentially eliminates
their rotation.  The solidified spheres no longer form chains, but
instead organize themselves into levitated colloidal crystals.  Given
that polymerization changes nothing about the spheres except for their
viscoelastic properties, we propose that the fluid droplets spin
because they are deformable and that their spinning gives rise to
hydrodynamic interactions that foster chain formation.  This mechanism
for sound-mediated self-organization appears not to have been reported
previously.

Section~\ref{sec:experiment} presents the model experimental system
that exhibits acoustically-driven spinning and chaining.
Section~\ref{sec:acoustohydrodynamicforces} explains how hydrodynamic
interactions among spinning spheres can induce chaining, particularly
when the spheres inherently are attracted to each other.  This
analysis reveals a threshold spinning rate above which planar crystals
are unstable and single-file chains are favored.  Simulations
described in Sec.~\ref{sec:simulation} then show that chain formation
proceeds at a rate consistent with experimental observations given the
inferred rate of single-sphere spinning.

\section{Experimental Observations}
\label{sec:experiment}

\subsection{Acoustic levitation}
Experiments are carried out in a cylindrical ultrasonic resonator made
of aluminium with inner diameter $D = \SI{20}{\milli\meter}$ and
height $H = \SI{375}{\micro\meter}$.  The top of the cavity is sealed
with a round quartz cover plate that serves as the reflector shown in
Fig.~\ref{fig:phenomenology}(a).  The transparent cover also provides
optical access to the sample.  The bottom is a \SI{0.30}{\mm}-thick
silicon wafer that launches sound waves into the sample.  A standing
wave is excited in this resonator by a \SI{2}{\mega\hertz}
piezoeletric transducer that is glued directly to the silicon wafer.
The disk-like transducer is centered on the wafer to maintain the
cylindrical symmetry of the resonator.  The transducer is driven
sinusoidally by a signal generator (TiePie HandyScope HS5) at
amplitudes up to \SI{12}{\volt} peak to peak
\cite{dron2012acousticenergy}.  The driving frequency is tuned to the
fundamental mode of the cavity with wavelength $\lambda = 2H$ as
depicted in Fig.~\ref{fig:phenomenology}(b).  For the experiments
reported here, the optimal driving frequency is
$\omega = \SI{1.985(1)}{\mega\hertz}$, which corresponds to a
wavelength of $\lambda \approx \SI{750}{\um}$ in water.

\subsection{TPM Droplets}
\label{sec:fluiddroplets}

An emulsion of monodisperse fluid droplets is prepared by homogeneous
nucleation of 3-(trimethoxysilyl) propyl methyl methacrylate (TPM)
($\ge \SI{98}{\percent}$, Sigma Aldrich) in a reducing environment
following the procedure described in \cite{vanderwel2017preparation}.
All chemicals are used as received with no further purification.  The
emulsification medium is composed of ammonia (\ce{NH3},
\SI{28}{wt.\percent}, Sigma Aldrich) diluted 1000:1 in deionized water
(\SI{18.2}{\mega\ohm\cm}).  Droplets are formed by adding
\SI{1}{\milli\liter} of TPM to \SI{100}{\milli\liter} of the medium at
room temperature with gentle stirring for \SI{20}{\minute}.  The
diameter of the droplets is then increased by adding more TPM.  Four
additions of \SI{600}{\micro\liter} of TPM at \SI{20}{\minute}
intervals increases the final droplet radius to $a_p = \SI{1.8}{\um}$
with a polydispersity in radius of \SI{2.2}{\percent}.  The size and
polydispersity of the TPM droplets are measured by holographic
particle characterization (Spheryx xSight) and confirmed by dynamic
light scattering (Malvern ZetaSizer Nano ZS).  The droplets are dyed
with FITC to facilitate fluorescence microscopy
\cite{vanderwel2017preparation}.

A small amount of an oil-soluble free-radical photoinitiator
($\SI{1}{wt\percent}$, Duracure 1173, ciba) is added to the droplets
once growth is complete.  The photoinitiator is used to polymerize the
fluid droplets into solid spheres, but has no influence on their
properties until triggered.  Completed droplets are transferred to
deionized water and are electrostatically stabilized with the addition
of \SI{1}{\milli M} sodium dodecyl sulfate (SDS). Holographic
characterization measurements confirm that the droplets remain stable
and monodisperse for at least 5 weeks.

TPM droplets condensed at low ammonia concentration have a mass
density of $\rho_p = \SI{1.18(1)}{\gram\per\cubic\cm}$, as determined
by sedimentation equilibrium in a sugar gradient.  This is
significantly lower than the value of
\SI{1.235(10)}{\gram\per\cubic\cm} obtained in more strongly reducing
media \cite{vanderwel2017preparation}.  These droplets have a
correspondingly low refractive index of $n_p = \num{1.482(3)}$ at a
vacuum wavelength of \SI{447}{\nm} as determined by holographic
characterization, which is significantly smaller than the value of
\num{1.506(7)} reported for solid TPM spheres
\cite{middleton2019optimizing}.  The density contrast between the
droplets and their aqueous environment is large enough to facilitate
acoustic trapping but not large enough for gravity to compete
effectively with acoustic forces at experimentally accessible pressure
levels.

\subsection{Solid TPM spheres}
\label{sec:solidspheres}

TPM droplets are solidified by exposing them to ultraviolet radiation
(\SI{365}{\nm}, $\SI{3.5}{\milli\watt\per\square\cm}$) for
\SI{15}{\minute} to trigger free-radical polymerization.  The
polymerized TPM spheres have the same mean diameter and polydispersity
as their fluid progenitors.  Their density, however, increases by
\SI{3}{\percent} to $\rho_p = \SI{1.22(1)}{\gram\per\cubic\cm}$.  The
increase in density is consistent with the observed increase in
refractive index to $n_p = \num{1.508(2)}$ and also is consistent with
results of previous characterization studies
\cite{vanderwel2017preparation,middleton2019optimizing}.
Photopolymerized spheres are washed three times and redispersed in
\SI{1}{\milli M} SDS solution.

\subsection{Colloidal Imaging}
\label{sec:imaging}

Bright-field and fluorescence images are captured in a Nikon Eclipse
Ti widefield fluorescence microscope in reflection mode.  The
$60\times$ oil-immersion objective provides a system magnification of
\SI{120}{\micro\meter\per\pixel} and the camera records
$\num{1280} \times \num{1280}$ pixel images at
\SI{30}{frames\per\second} with an exposure time of
\SI{100}{\micro\second}.

\subsection{Formation of crystalline monolayers}

A dispersion of solid colloidal TPM spheres is driven to the midplane
of the ultrasonic levitator in a matter of seconds and forms monolayer
crystals in a few minutes.  The spheres within a crystal appear to be
in contact, as can be seen in the inset to
Fig.~\ref{fig:phenomenology}(c) and individual spheres do not appear
to move relative to their neighbors once the crystal has formed.
Levitated crystals are stable and drift freely in the nodal plane.
Crystallization is reversible; turning off the ultrasound frees the
spheres to diffuse and sediment independently.

\subsection{Chaining of fluid droplets}
\label{sec:chaining}

Performing the same levitation experiments with droplets rather than
rigid spheres yields substantially different behavior.  Individual
droplets appear to vibrate rapidly in the imaging plane, which we
interpret to be the projection of rapid rotation around an axis
parallel to the imaging plane. Rather than packing into
two-dimensional crystals, these droplets form chains that are aligned
with the presumed axis of rotation and perpendicular to the axis of
the observed vibration.  As for the solid-sphere crystals, the
droplets within a chain appear to be in contact with each other.
Droplet chaining is reversible, and the droplets diffuse apart and
sediment as soon as the levitator is turned off.

\begin{figure}
  \centering \includegraphics[width=0.7\columnwidth]{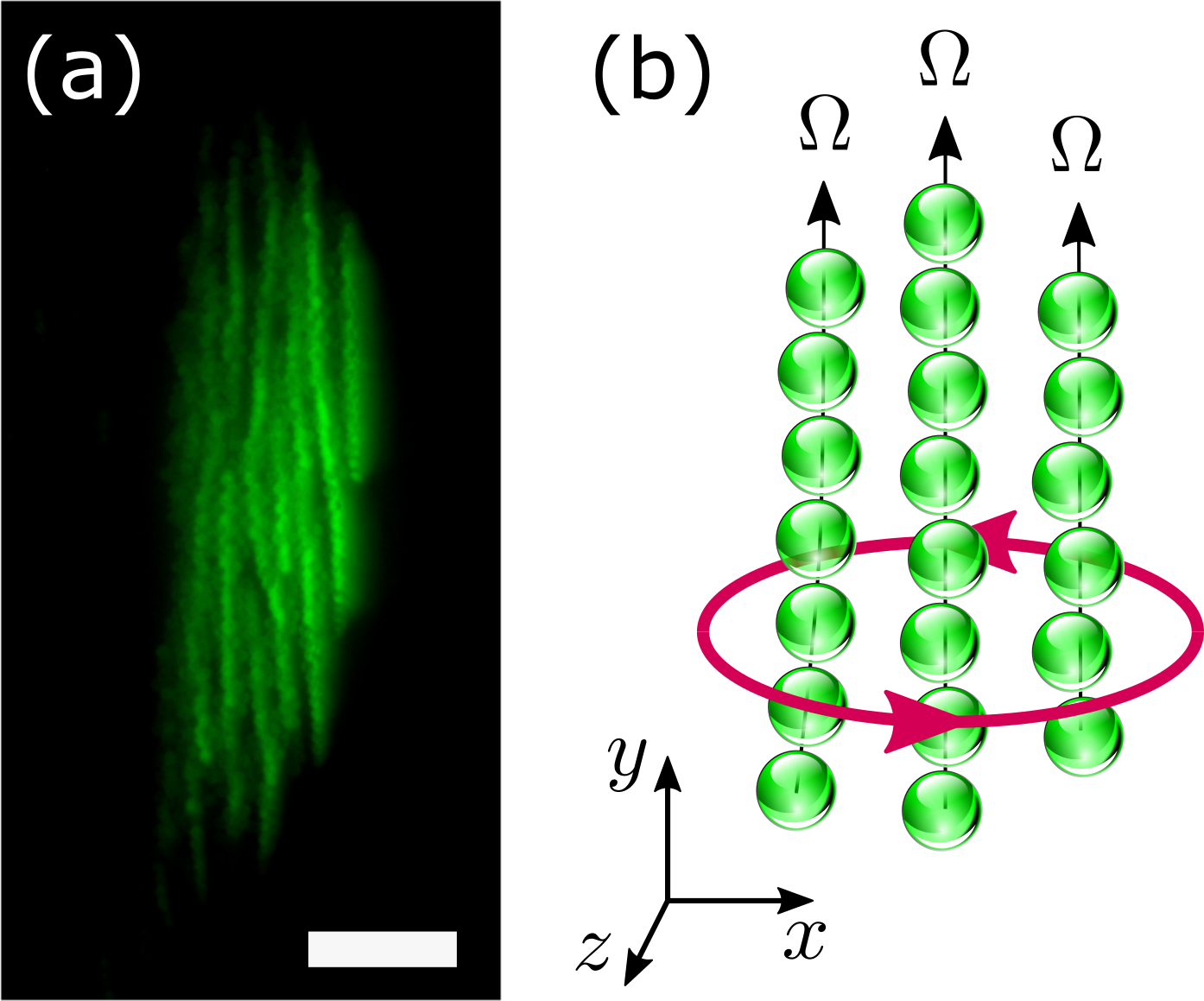}
  \caption{Chains of rapidly rotating emulsion droplets form bundles
    that slowly orbit each other. (a) Fluorescence microscopy image of
    acoustically levitated TPM droplets forming a bundle of
    chains. Scale bar represents \SI{50}{\um}.  (b) Schematic
    representation of droplets spinning with angular velocity
    $\vec{\Omega}$ forming a bundle of regularly spaced chains that
    collectively rotate.}
  \label{fig:chainbundles}
\end{figure}

Depending on initial conditions, chains either form in comparative
isolation, as in Fig.~\ref{fig:phenomenology}(d), or they form bundles
such as the example in Fig.~\ref{fig:chainbundles}(a).  The chains
within a bundle appear to maintain uniform spacing from each other.
They do not appear to exchange droplets.  Bundled chains, moreover,
tend to orbit each other, as indicated schematically in
Fig.~\ref{fig:chainbundles}(b).  This orbital motion can be explained
naturally if the individual spheres are spinning because the resulting
rotational flows would advect neighboring chains.  Individual droplets
also are observed to orbit nearby chains.

The bundles' three-dimensional orbital motion carries the spinning
droplets tens of micrometers above and below the stable trapping plane
near the node of the ultrasonic pressure field.  This out-of-plane
motion contrasts with the rigid planarity of solid-sphere crystals.

These observations inspire our proposal that individual droplets are
driven into rapid rotation by the ultrasonic levitator and that the
resulting hydrodynamic coupling among neighboring spheres disrupts
crystallization and instead organizes the droplets into chains.  The
remaining role for acoustic forces is to draw the chained droplets
into contact.  Similar rotation and chaining has been reported in
acoustic levitation experiments on metallic \cite{zhou2017twists} and
bimetallic nanorods
\cite{wang2012autonomous,balk2014kilohertz,nadal2014asymmetric,lippera2019nonetmotion,dumy2020acoustic}.
Zhou, \emph{et al.} also report sound-induced rotation of
silica-titanium Janus microspheres \cite{zhou2017twists}.  The
particles in all of these previous reports have lower symmetry than
the homogeneous spherical particles and droplets in our experiments.
Whereas fluid droplets normally are spherical, they also are
deformable and their deformability may contribute to their propensity
to spin.

\section{Acoustohydrodynamic Forces}
\label{sec:acoustohydrodynamicforces}

\subsection{Acoustic Forces}
\label{sec:acousticforces}

Sound transports objects by advecting them in its velocity field and
by exerting forces on them through pressure gradients.  In a nearly
incompressible medium such as water, a sound wave's pressure,
$p(\vec{r}, t)$, acts as a scalar potential for the velocity,
\begin{equation}
  \vec{v}(\vec{r}, t) = -\frac{i}{\omega \rho_m} \nabla p ,
\end{equation}
where $\omega$ is the wave's angular frequency and $\rho_m$ is the
mass density of the medium.  In this case, the leading-order
contributions to the time-averaged force exerted by a harmonic sound
wave on a small object may be expressed in terms of the pressure alone
as \cite{silva2014acoustic,abdelaziz2020acoustokinetics}
\begin{subequations}
  \label{eq:acousticforces}
  \begin{equation}
    \vec{F}_a(\vec{r})
    =
    \frac{1}{2} \real{
      \alpha \, p \nabla p^*
      +
      \beta \, k^{-2} (\nabla p \cdot \nabla) \nabla p^*},
    \label{eq:forceprinciple}
  \end{equation}
  where $k = \omega/c_m$ is the wavenumber of the sound in a medium
  with sound speed $c_m$.

  The acoustic force described by Eq.~\eqref{eq:forceprinciple}
  depends on the object's dipole and quadrupole acoustic
  polarizabilities, $\alpha$ and $\beta$, respectively.  These, in
  turn, depend on the object's radius, $a_p$, density, $\rho_p$, and
  sound speed, $c_p$:
  \begin{align}
    \alpha
    & =
      \frac{4 \pi a_p^3}{3 \rho_m c_m^2}
      f_0
      \left[
      -1 + \frac{i}{3} (f_0 + f_1) (k a_p)^3
      \right] \label{eq:alpha} \\
    \beta
    & =
      \frac{2 \pi a_p^3}{\rho_m c_m^2}
      f_1
      \left[
      1 + \frac{i}{6} f_1 (ka_p)^3
      \right].
      \label{eq:beta}
  \end{align}
  Within these expressions, the monopole coupling coefficient,
  \begin{equation}
    f_0 = 1 - \frac{\rho_m c_m^2}{\rho_p c_p^2},
    \label{eq:f_0}
  \end{equation}
  depends on the compressibility of the particle,
  $\kappa_p = 1/(\rho_p c_p^2)$, relative to that of the medium, and
  the dipole coupling coefficient depends on the density mismatch,
  \begin{equation}
    f_1 = 2 \frac{\rho_p - \rho_m}{2 \rho_p + \rho_m}.
    \label{eq:f_1}
  \end{equation}
\end{subequations}
For TPM in water, $f_0 = \num{-0.33}$ and $f_1 = \num{0.13}$, assuming
$c_p = \SI{1040}{\meter\per\second}$ \cite{xu2020acoustic}.
Equation~\eqref{eq:acousticforces} should accurately predict acoustic
forces experienced by objects that are smaller than the wavelength of
sound, $k a_p < 1$.  The water-borne TPM droplets and spheres
described in Sec.~\ref{sec:experiment} have a reduced size of
$k a_p = \num{0.03}$ when levitated at \SI{2}{\mega\hertz} and so
satisfy this condition.

\subsubsection{Acoustic levitation}
\label{sec:acousticlevitation}

The acoustic levitator used in this study is a resonant cavity whose
height, $H$, is adjusted to half a wavelength, $kH = \pi$, creating a
standing pressure wave with a node along the midplane:
\begin{equation}
  p_0(\vec{r}, t)
  =
  p_0 \, \sin(k z) \, e^{-i\omega t}.
  \label{eq:standingwave}
\end{equation}
This sound wave exerts a primary acoustic force,
\begin{align}
  \vec{F}_1(z)
  & =
    \frac{1}{4} k \, p_0^2 \, (\alpha' - \beta') \,
    \sin(2 k z) \, \hat{z},
    \label{eq:acoustictrap} \\
  & \approx
    - \frac{1}{2} k^2 p_0^2 \, (\beta' - \alpha') \, z \,
    \hat{z} ,
    \label{eq:hookean}
\end{align}
that is directed vertically along $\hat{z}$.  Primes in
Eq.~\eqref{eq:acoustictrap} denote the real parts of the
polarizabilities.  Equations~\eqref{eq:alpha} through \eqref{eq:f_1}
show that $\beta' > \alpha'$ for dense incompressible objects.  The
levitator therefore localizes such objects near the midplane at
$z = 0$ with an approximately Hookean restoring force that is
proportional to the sound wave's intensity.

The energy density stored in the standing wave reaches
$\epsilon_0 = \SI{20}{\joule\per\cubic\meter}$ at the maximum driving
voltage.  This sets the scale for the standing wave's amplitude
through
\begin{equation}
  \epsilon_0 = \frac{1}{2} \frac{p_0^2}{\rho_m c_m^2}.
\end{equation}

\subsubsection{Secondary Bjerknes interaction}
\label{sec:secondarybjerknes}

Objects trapped in the levitator interact with each other via
scattered sound waves, a mechanism known as the secondary Bjerknes
interaction.  The nature of this interparticle coupling also may be
elucidated with Eq.~\eqref{eq:acousticforces}.  To leading order in
$k a_p$, a small object located at $\vec{r}_j$ scatters spherical
waves to its neighbor at $\vec{r}_i$ of the form
\cite{silva2014acoustic}
\begin{subequations}
  \label{eq:scatteredpressure}
  \begin{equation}
    p_s(\vec{r}_i, \vec{r}_j, t)
    =
    \left[
      - \Phi_p \,
      + \Phi_v \nabla_i \cdot \nabla_j
    \right]
    \frac{e^{i k r_{ij}}}{k r_{ij}}
    p_0(\vec{r}_j,t),
    \label{eq:sphericalpressurefield}
  \end{equation}
  where $\vec{r}_{ij} = \vec{r}_i - \vec{r}_j$.  The scattered
  pressure field includes a contribution proportional to the pressure
  of the standing wave at the position of the scatterer with
  coefficient
  \begin{equation}
    \Phi_p = \frac{1}{3} (k a_p)^3 \, f_0
  \end{equation}
  and another term proportional to the standing wave's velocity with
  coefficient
  \begin{equation}
    \Phi_v = \frac{1}{2} (k a_p)^3 \, f_1.
  \end{equation}
\end{subequations}
The next-higher-order contributions to $p_s(\vec{r}_i, \vec{r}_j, t)$
have gradients directed along $\hat{z}$ and so do not contribute to
interparticle interactions.

A particle located at $\vec{r}_i$ experiences both the levitator's
pressure field and also the scattered field due to its neighbors.
Considering just one such interaction, the net pressure wave
experienced by a particle at $\vec{r}_i$ is
\begin{equation}
  p(\vec{r}_i, \vec{r}_j, t)
  = p_0(\vec{r}_i, t) + p_s(\vec{r}_i, \vec{r}_j, t),
\end{equation}
which gives rise to a total acoustic force,
\begin{equation}
  \vec{F}_a(\vec{r}_i, \vec{r}_j) =
  \vec{F}_1(z_i) + \vec{F}_2(\vec{r}_i, \vec{r}_j),
\end{equation}
that combines the influence of the acoustic trap and the secondary
Bjerknes interaction.

\begin{figure}
  \centering \includegraphics[width=0.5\columnwidth]{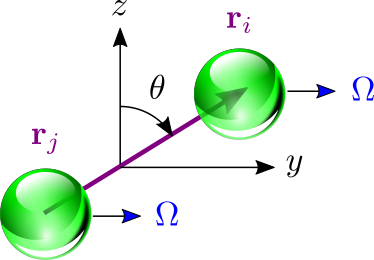}
  \caption{Geometry of acoustohydrodynamic forces acting on
    acoustically levitated spheres.  The sphere at $\vec{r}_i$ is
    drawn to the nodal plane at $z = 0$ by the primary acoustic
    radiation force, $\vec{F}_1(\vec{r}_i)$, and experiences a
    secondary Bjerknes force, $\vec{F}_2(\vec{r}_i, \vec{r}_j)$, due
    to its neighbor at $\vec{r}_j$.  The spheres' separation is
    inclined at angle $\theta$ with respect to the vertical $\hat{z}$
    axis. Both spheres are assumed to be spinning with angular
    velocity $\vec{\Omega}$, directed along $\hat{y}$.}
  \label{fig:geometry}
\end{figure}

Particles trapped near the pressure node at $z = 0$ experience
$p_0(\vec{r}, t) \approx 0$ and
$\nabla p_0 \approx k p_0 \exp(-i \omega t) \hat{z}$.  Their secondary
Bjerknes interaction therefore is dominated by the velocity-dependent
part of Eq.~\eqref{eq:scatteredpressure}.  This is still the case if
the particles are displaced slightly from $z = 0$ by gravity.
Assuming that the trapped particles all have the same physical
properties, their secondary Bjerknes interaction reduces to a
conservative pairwise-additive force,
\begin{multline}
  \vec{F}_2(\vec{r}_i, \vec{r}_j) =
  \vec{F}_2(r_{ij}, \theta) \\
  = \frac{3}{2} k p_0^2 \frac{\beta' \Phi_v}{(kr_{ij})^4} \left[
    \frac{1+3 \cos (2\theta)}{2} \, \hat{r} +
    \sin(2\theta) \, \hat{\theta} \right] \\
  + \mathcal{O}\left\{ (kr_{ij})^{-3}, (ka_p)^9 \right\} ,
  \label{eq:secondarybjerknes}
\end{multline}
where $\theta$ is the angle between $\vec{r}_{ij}$ and $\hat{z}$, as
shown in Fig.~\ref{fig:geometry}.
Equation~\eqref{eq:secondarybjerknes} further reduces to the classic
expression for the secondary Bjerknes interaction
\cite{silva2014acoustic} when the two spheres are in the nodal plane,
$\theta = \pi/2$.  This in-plane interaction is isotropically
attractive and tends to organize monodisperse spheres into
close-packed crystals such as the example in
Fig.~\ref{fig:phenomenology}(c).  The interaction becomes repulsive if
the particles are canted by less than $\theta = \SI{35.3}{\degree}$
from the vertical.  This sign change influences sound-mediated
organization when the particles are displaced from the trapping plane
by competing influences such as hydrodynamic forces.

Interestingly, $\vec{F}_2(r_{ij}, \theta)$ depends on the particles'
density through $\beta'$ and $\Phi_v$, but not on their
compressibility.  The secondary Bjerknes interaction therefore does
not distinguish between fluid droplets and solid spheres of the same
size and density.  Differences in their phenomenology therefore must
be due to other mechanisms.

\subsection{Hydrodynamic forces}
\label{sec:hydrodynamicforces}

If we accept the proposal that fluid droplets tend to spin rapidly
around a common axis in the trapping plane, then their rotation drives
circulatory fluid flows that mediate hydrodynamic interactions.
Although a definitive explanation for the observed spinning is not yet
available, Appendix~\ref{sec:torque} presents one possible mechanism.
The far-field flow generated by a sphere of radius $a_p$ located at
$\vec{r}_j$ and rotating around the $\hat{y}$ axis at frequency
$\Omega$ may be modeled as a rotlet,
\begin{equation}
  \vec u(\vec{r} - \vec{r}_j)
  =
  \Omega \frac{a_p^3}{\abs{\vec{r} - \vec{r}_j}^2}
  \sin \theta \, \hat{\phi},
  \label{eq:rotlet}
\end{equation}
where $\theta$ is the polar angle indicated in Fig.~\ref{fig:geometry}
and $\phi$ is the azimuthal angle around $\hat{z}$.
Equation~\eqref{eq:rotlet} is a simplifying approximation for the flow
field due to a spinning droplet because it does not account for
fluid-fluid boundary conditions \cite{leal2007advanced} or for the
droplet's deformations.  We also neglect the influence of bounding
walls on $\vec{u}(\vec{r} - \vec{r}_j)$ because the inter-particle
separation is much smaller than the distance to the nearest wall.

The flow described by Eq.~\eqref{eq:rotlet} exerts a hydrodynamic
force on a neighboring sphere at $\vec{r}_i$ that is described by
Fax\'en's first law:
\begin{equation}
  \vec F_h(\vec{r}_i, \vec{r}_j)
  =
  \gamma
  \left(1 + \frac{a_p^2}{6} \nabla_i^2 \right)
  \vec{u}(\vec{r}_i - \vec{r}_j),
\end{equation}
where $\gamma = 6 \pi \eta_m a_p$ is the Stokes drag coefficient in a
medium of viscosity $\eta_m$.  The Laplacian term vanishes in the
rotational flow and sphere $i$ is simply advected by its neighbor's
flow field.

\subsection{Acoustohydrodynamic chaining}
\label{sec:acoustohydrochaining}

The chaining mechanism can be understood by considering the dynamics
of two acoustically levitated spheres undergoing coaxial rotation in
the trapping plane.  By Fax\'en's law, the spheres' center of mass,
$\vec{R} = (\vec{r}_1 + \vec{r}_2)/2$, moves as
\begin{multline}
  2 \gamma \dot{\vec{R}} = \vec{F}_a(\vec{r}_1, \vec{r}_2)
  + \vec{F}_a(\vec{r}_2, \vec{r}_1) \\
  + \vec{F}_h(\vec{r}_1, \vec{r}_2) + \vec{F}_h(\vec{r}_2, \vec{r}_1)
  + 2 \vec{F}_g + 2 \vec{F}_1(\vec{R}),
  \label{eq:centerofmass}
\end{multline}
where the dot denotes a derivative with respect to time and
$\vec{F}_g$ is the force of gravity acting on the spheres' buoyant
masses.  We assume that $\vec{F}_g$ is weak enough for the acoustic
trapping force to be linear in vertical displacements, as described by
Eq.~\eqref{eq:hookean}.  In that case,
$\vec{F}_1(\vec{r}_1) + \vec{F}_1(\vec{r}_2) \approx 2
\vec{F}_1(\vec{R})$.  We further assume that the spheres'
rotation-induced interaction is reciprocal:
$\vec{F}_h(\vec{r}_1, \vec{r}_2) = - \vec{F}_h(\vec{r}_2, \vec{r}_1)$.

The spheres' separation, $\vec{r} = \vec{r}_1 - \vec{r}_2$, evolves as
\begin{multline}
  \label{eq:polarAngle}
  \gamma \dot{\vec{r}} = \vec{F}_a(\vec{r}_1, \vec{r}_2)
  - \vec{F}_a(\vec{r}_2, \vec{r}_1) \\
  + \vec{F}_h(\vec{r}_1, \vec{r}_2) - \vec{F}_h(\vec{r}_2, \vec{r}_1)
  +\vec{F}_1(\vec{r}).
\end{multline}
The secondary Bjerknes interaction tends to draw the spheres into
contact.  We assume therefore that the relative separation does not
not vary significantly, $\dot{r} \approx 0$.  In spherical
coordinates, $\vec{r} = (r, \theta, \phi)$, the pair's azimuthal
orientation then evolves as
\begin{equation}
  \dot{\phi}
  =
  2 \Omega \, \frac{a_p^3}{r^3} \,
  \cos \theta \, \sin \phi,
  \label{eq:azimuthal}
\end{equation}
which would describe uniform rotation about $\hat{y}$ if the polar
angle, $\theta$, were fixed.  The spheres' polar orientation also can
evolve in time, however, and the full expression for $\dot{\theta}$ is
presented in Appendix~\ref{sec:polarAngle}.

Equation~\eqref{eq:azimuthal} has a fixed point, $\dot{\phi} = 0$, for
$\theta = \pi/2$, which corresponds to both spheres lying in the
horizontal plane at the equilibrium height set by
Eq.~\eqref{eq:centerofmass}.  In this configuration, the spheres'
polar orientation evolves as
\begin{equation}
  \dot{\theta}
  =
  2 \Omega \, \frac{a_p^3}{r^3} \, \cos\phi.
\end{equation}
The stable configuration with $\phi = \pi/2$ (or equivalently
$\phi = 3 \pi/2$) corresponds to the spheres aligning along their
common rotation axis, $\hat{y}$.  Chaining therefore represents a
stable fixed point for the configuration of coaxially spinning
particles.

Alternatively, Eq.~\eqref{eq:azimuthal} reveals that stable
configurations can appear at $\phi = 0$, with the spheres separated
along $\hat{x}$, perpendicular to their rotation axis, $\hat{y}$.  The
radial component of the hydrodynamic force is repulsive in this
configuration \cite{jeffery1915steady}.  Stable transverse
configurations therefore require an independent compensating
attraction, such as the secondary Bjerknes force.

Mechanically stable solutions with $\phi = 0$ and $\dot{\theta} = 0$
are impossible if the spheres' rotation rate exceeds a critical value,
\begin{equation}
  \label{eq:Omega_c}
  \Omega_c
  \approx
  \frac{p_0^2}{\eta_m \rho_m c_m^2}
  \left[
    \frac{f_1^2}{16}
    +
    \frac{2f_0 + 3f_1}{9} (ka_p)^2
  \right].
\end{equation}
A transverse pair of rapidly spinning spheres tumbles out of the
trapping plane, orbiting the center of mass indefinitely at angular
frequency $\dot{\theta} = 2\Omega$.

Slowly spinning spheres with $0 < \Omega < \Omega_c$ can be stably
oriented at
\begin{equation}
  \theta
  =
  \pi
  - \frac{1}{2} \,
  \arcsin\bigg( \frac{\Omega}{\Omega_c} \bigg)
\end{equation}
with respect to the vertical axis.  This reflects a mechanical
equilibrium between axial acoustic forces and rotation-induced
hydrodynamic forces.  The non-spinning limit at $\Omega = 0$ has a
fixed point at $\theta = \pi/2$, which corresponds to conventional
in-plane self-organization mediated by the secondary Bjerknes
interaction.  An additional fixed point at $\theta = 0$ has the
spheres stacked atop each other along the vertical axis, but is
unstable.  At $\Omega = \Omega_c$, the marginally stable transverse
configuration is inclined at $\theta = 3\pi/4$.

The tumbling solution with $\Omega > \Omega_c$ remains stationary in
the sense that $\dot{\phi} = 0$ only if $\phi \approx \pi/2$.
Deviations from transversality allow a tumbling pair to evolve into
the aligned state at $\phi = 0$ and therefore to form a chain.  This
analysis suggests that the combination of acoustic and hydrodynamic
forces tends to align pairs of co-rotating spheres along their common
axis of rotation provided that their rotation rate is high enough,
$\Omega > \Omega_c$.  For \SI{3.6}{\um}-diameter TPM spheres trapped
in our levitator, the critical rotation rate is
\begin{equation}
  \Omega_c \approx \frac{1}{8} \frac{\epsilon_0}{\eta_m} \, f_1^2
  \approx \SI{7}{\hertz}.
\end{equation}
The tendency of co-rotating pairs to align along their common axis of
rotation inspires us to propose that hydrodynamic forces tend to
organize larger ensembles of rapidly-rotating spheres into linear
chains while acoustic forces shepherd more slowly rotating particles
into planar crystals.

\section{Numerical simulations of chain formation}
\label{sec:simulation}

We test the prediction that rotation-mediated acoustohydrodynamic
forces create chains through molecular dynamics simulation
\cite{frenkel2001understanding}.  The total force acting on the $i$-th
sphere in an $N$-sphere cluster is
\begin{equation}
  \vec{F}_i(\vec{r}_i)
  =
  \vec{F}_g
  + \sum_{j \neq i}
  \vec{F}_h(\vec{r}_i, \vec{r}_j)
  +
  \vec{F}_a(\vec{r}_i, \vec{r}_j) .
\end{equation}
That sphere's velocity then depends on all of the forces acting on the
system,
\begin{subequations}
  \begin{equation}
    \dot{\vec{r}}_i
    =
    \gamma^{-1} \sum_{j = 1}^N G_{ij} \vec{F}_j(\vec{r}_j),
  \end{equation}
  through the Oseen tensor \cite{happel2012low},
  \begin{equation}
    G_{ij}
    =
    \frac{3a}{4}
    \left(
      \frac{\delta_{ij}}{r}
      + \frac{\vec{r}_{ij} \vec r_{ij}}{r^3}\right)
    + \frac{a^3}{4}
    \left(
      \frac{\delta_{ij}}{r^3}
      - 3 \frac{\vec{r}_{ij} \vec{r}_{ij}}{r^5} \right),
  \end{equation}
  that describes hydrodynamic coupling among the spheres.
  \label{eq:motion}
\end{subequations}

Equation~\eqref{eq:motion} incorporates several simplifying
assumptions.  It ignores inertial effects under the assumption that
the spheres' motions are overdamped.  It also neglects diffusion under
the assumption that thermal forces are much weaker than either
acoustic or hydrodynamic forces.  The forces themselves are treated in
their leading-order approximations.  Leading-order expressions for
acoustic forces omit non-conservative and non-additive interactions.
The leading-order Oseen tensor does not account for near-field
hydrodynamic interactions or lubrication forces.  Finally, we assume
for simplicity that the spheres all rotate at a fixed rate, $\Omega$,
independent of their configuration.  Invoking these approximations
yields a minimal model for chaining because it is unlikely that any of
the omitted effects would tend to promote chain formation over
crystallization.

The spheres' trajectories are calculated by numerically integrating
Eq.~\eqref{eq:motion} with the Euler method.  All forces and
velocities are recalculated at each time step.  Excluded volume
interactions are incorporated by backing particles away from
collisions \cite{frenkel2001understanding}.  Material parameters are
chosen to mimic the experimental system, with rotation rates ranging
from $\Omega = 0$ to $\Omega = \SI{100}{\hertz}$.  The 10 particles
initially are randomly distributed in a \SI{16}{\micro\meter} cube
centered at the coordinate origin, and then are allowed to reorganize
themselves under the influence of acoustic, hydrodynamic and
gravitational forces.

\begin{figure*}
  \centering \includegraphics[width=0.9\textwidth]{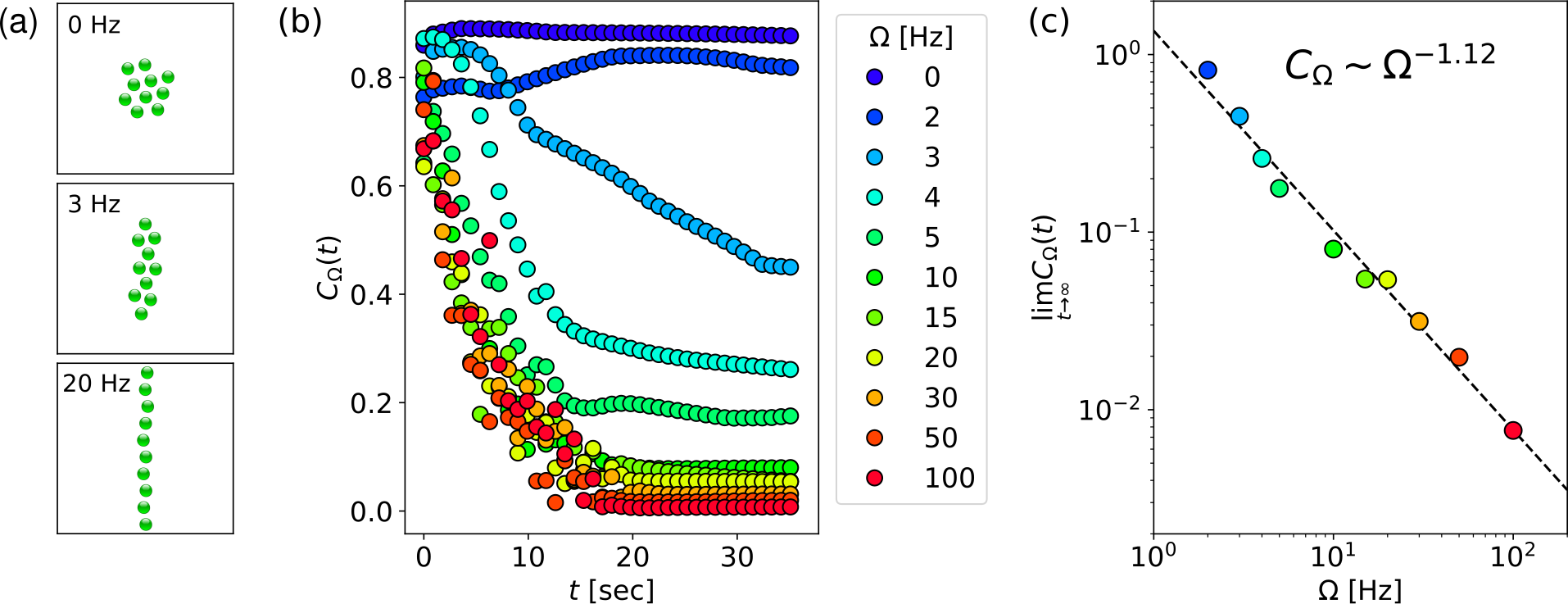}
  \caption{(a) Typical steady-state configurations for 10-particle
    clusters for three representative rotation rates: crystalline
    cluster at $\Omega = \SI{0}{\hertz}$ single-file chain at
    $\Omega = \SI{20}{\hertz}$ and intermediate state at
    $\Omega = \SI{3}{\hertz}$.  (b) Influence of the rotation rate,
    $\Omega$, on chain formation as monitored by the time evolution of
    the computed cluster compactness, $C_\Omega(t)$.  (c) The
    converged compactness, $\lim_{t \to \infty} C_\Omega(t)$, scales
    with rotation rate, which suggests that crystals evolve
    continuously into chains as the rotation rate increases.  The
    dashed line is a fit to the scaling form
    $C_\Omega \propto \omega^{-\nu}$ with scaling exponent
    $\nu = \num{1.12}$.}
  \label{fig:sim}
\end{figure*}

Particles move rapidly toward the trapping plane in these simulations
and then reorganize themselves into configurations such as the
examples in Fig.~\ref{fig:sim}(a).  The set of single-particle
trajectories, $\{\vec{r}_i(t)\}$, can be compared both with
predictions of the analytical theory for pair dynamics and also with
experimental observations on large collections of spheres.  Examples
of animated trajectories for slow and fast rotation are presented in
the Supplemental Material\footnote{See Supplemental Material at [URL
  will be inserted by publisher] for video animations of molecular
  dynamics simulations.}.

We quantify the degree of chaining by computing the projection of the
particles' convex hull in the horizontal plane
\cite{preparata2012computational}.  The area of the polygonal convex
hull, $A(\{\vec{r}_i\})$, vanishes for an ideal chain and reaches a
maximal value for a compact crystalline cluster.  We therefore define
a measure of compactness for a system with rotation rate $\Omega$,
\begin{equation}
  C_\Omega(t)
  =
  4\pi \frac{A(\{\vec{r}_i(t)\})}{P^2(\{\vec{r}_i(t)\})} ,
\end{equation}
where $P(\{\vec{r}_i)\})$ is the perimeter of the convex hull.  This
metric approaches $C_\Omega(t) = 1$ for a large crystal and vanishes
for a perfectly aligned chain.

Figure~\ref{fig:sim}(b) shows typical examples of the time evolution
of $C_\Omega(t)$ for $N = 10$ at different values of $\Omega$.  As
expected, non-rotating spheres form compact clusters while rapidly
rotating spheres approach $\lim_{t \to \infty} C_\Omega(t) = 0$.  The
transition between these two limiting behaviors appears to be
continuous and is consistent with the estimated value of the critical
rotation rate, $\Omega_c \approx \SI{7}{\hertz}$.

The nature of the crystal-to-chain transition is clarified in
Fig.~\ref{fig:sim}(c), which shows the long-time asymptotic behavior
of $C_\Omega(t)$ as a function of rotation rate.  The compactness
metric scales as $\Omega^{-\nu}$ over two orders of magnitude with
exponent $\nu = 1.12$.  This suggests that crystals extend
continuously into chains as the spheres spin faster.  A typical
example of the intermediate state at $\Omega = \SI{3}{\hertz}$ appears
in Fig.~\ref{fig:sim}(a).

These numerical studies confirm that the acoustohydrodynamic model
accounts for the observed chaining of TPM fluid droplets in good
quantitative agreement with experiment.  This mechanism requires the
fluid droplets to spin along a common axis at a rate that exceeds a
critical value $\Omega > \Omega_c$.  Solid spheres presumably rotate
less rapidly and therefore form conventional compact crystals.
Neither the scaling form for the crystal-chain transition nor the
observed exponent are yet explained.

\section{Conclusions}
\label{sec:conclusions}

We have demonstrated experimentally, theoretically and in simulations
that spinning spheres tend to form chains when levitated by an
acoustic standing wave.  Hydrodynamic forces engendered by coaxial
spinning favor alignment along the common axis.  They do not, however,
provide the pair attraction required to coalesce clusters of spheres
into chains.  This attraction is provided in our system by the
secondary Bjerknes interaction, which operates independently of the
spheres' spinning.

Spin-mediated hydrodynamic coupling competes with the confining
potential of the acoustic levitator.  Slowly spinning spheres remain
trapped in the plane and so form crystals rather than reorganizing
themselves into chains.  We anticipate, therefore, that crystals of
slowly rotating spheres may only be metastable with respect to
chaining but that kinetic barriers to reorganization are prohibitive
unless the spheres' rotation rate is sufficiently high.

We propose that the spheres' in-plane rotation is driven by inevitable
imperfections in the acoustic trap.  Estimates based on the parameters
for our experimental system suggest that our solid spheres rotate more
slowly than the critical rate and therefore form crystals.  We propose
that fluid droplets of the same material rotate more rapidly because
of their deformability in the acoustic force field.  This proposal is
consistent with experimental observations of fluid droplets'
collective motions.  We do not, however, have a complete explanation
for this effect, and present it as a outstanding challenge.

Both the tendency of insonated droplets to rotate rapidly and the
tendency of co-rotating spheres to form chains appear to be novel
phenomena.  This mechanism for forming long colloidal chains may be
useful for assembling model colloidal polymers
\cite{mcmullen2018freely}, linear microrobots
\cite{agrawal2020scale,yang2020reconfigurable} and acoustically
actuated optical elements \cite{martin1999magnetic}.  More generally,
this study illustrates how the choice of material properties can be
used to influence the pathway for sound-mediated self-organization of
soft-matter systems.

\section*{Acknowledgments}
This work was supported primarily by the National Science Foundation
through award number DMR-2104837.  The Nikon Eclipse Ti microscope and
Spheryx xSight particle characterization instrument used for this
study were purchased as shared facilities by the NYU Materials
Research Science and Engineering Center with support from the NSF
under award award number DMR-1420073.  M.H.\ acknowledges support BDI
from CNES-CNRS and Aide \`a la Recherche Grant CNES-France.  J.A.D.\
acknowledges support from the Simons Foundation.

\appendix

\section{Acoustokinetic torque on a sphere}
\label{sec:torque}

If sound waves incident on or scattered by a particle carry angular
momentum, the particle can experience a torque that causes it to spin.
This occurs for anisotropic particles in uniform sound fields
\cite{nadal2014asymmetric}, and for isotropic spheres in nonuniform
fields \cite{zhang2011acoustic,toftul2019acoustic}.  To dipole order,
the time-averaged torque experienced by a sphere at $\vec{r}_i$ may be
expressed in terms of the sound wave's velocity field as
\cite{silva2014acoustic}
\begin{equation}
  \label{eq:torque}
  \vec{\tau}(\vec{r})
  =
  i \frac{12}{5}
  \frac{\rho_m^2 \rho_p}{(\rho_m + 2 \rho_p)^2}
  \frac{\omega}{c_p} \,
  \alpha_p(\omega) \,
  a_p^5
  \left(\vec{v} \times \vec{v}^*\right),
\end{equation}
where $\alpha_p(\omega)$ is the particle's acoustic attenuation
coefficient.  For a fluid droplet of viscosity $\eta_p$,
\begin{equation}
  \alpha_p(\omega)
  =
  \frac{2}{3} \frac{\eta_p}{\rho_p c_p^3} \,
  \omega^2.
\end{equation}
An ideally rigid sphere has $\alpha_p(\omega) = 0$ and so experiences
no torque and does not spin.  This distinction is consistent with our
observation that viscoelastic TPM droplets behave differently in an
acoustic levitator from solidified TPM spheres.

Expressing a general acoustic pressure field in terms of its
real-valued amplitude and phase,
\begin{equation}
  p(\vec{r}, t)
  =
  \abs{p(\vec{r})} e^{i \phi(\vec{r})} \,
  e^{-i \omega t},
\end{equation}
yields an expression for the torque in terms of the structure of the
field
\begin{equation}
  \vec{\tau}(\vec{r}) =
  i \frac{12}{5}
  \frac{\rho_p}{(\rho_m + 2 \rho_p)^2}
  \frac{1}{c_p \omega} \,
  a(\omega) \,
  a_p^5
  \left(
    \nabla \abs{p}^2 \times \nabla \phi
  \right)
  \label{eq:aktorque}
\end{equation}
that clarifies conditions under which acoustically levitated
viscoelastic spheres can spin.  An ideal standing wave has no phase
gradients, $\nabla \phi = 0$, and so exerts no torque.  A perfectly
uniform field with $\nabla \abs{p} = 0$ similarly exerts no torque.
Spheres only experience torques in nonuniform acoustic fields.

To illustrate how acoustic torque might arise in practice, we model
the wave launched by the piezoelectric transducer as a plane wave with
wave vector $\vec{k}$ and the reflected wave as a plane wave with wave
vector $\vec{q}$:
\begin{equation}
  p(\vec{r})
  =
  \frac{p_0}{\sqrt{2}}
  \left( e^{i \vec{k} \cdot \vec{r}}
    + e^{i \vec{q} \cdot \vec{r}} \right).
\end{equation}
Expressing $p(\vec{r})$ in terms of its amplitude and phase,
\begin{subequations}
  \begin{align}
    \abs{p(\vec{r})}
    & =
      p_0 \sqrt{1+\cos{(\vec{k} - \vec{q})
      \cdot \vec{r}}} \\
    \phi(\mathbf r)
    & =
      \frac{1}{2} (\vec{k} + \vec{q}) \cdot \vec{r},
  \end{align}
  then leads to
  \begin{equation}
    \nabla \abs{p}^2 \times \nabla \phi
    =
    p_0^2 \,
    \sin\left(\vec{k} \cdot \vec{r}
      - \vec{q} \cdot \vec{r}\right) \,
    \vec{k} \times \vec{q}.
  \end{equation}
  An ideal standing wave has $\vec{q} = -\vec{k}$ so that the sphere
  experiences no torque.  In practice, however, the incident and
  reflected waves may be misaligned by a small angle $\delta$.
  Orienting the coordinate system so that $\vec{k} = k \hat{z}$ and
  $\vec{q}$ is rotated about $\hat{x}$ yields
  \begin{equation}
    \vec{\tau}(\vec{r})
    =
    k^2 p_0^2 \, \sin \delta \,
    \sin(2 k z) \,
    \hat{y},
  \end{equation}
\end{subequations}
where $z$ is the particle's displacement from the antinode at $z = 0$.

To estimate the scale of the resulting rotation rate, we model a
droplet as a sphere with no-slip boundary conditions immersed in a
medium of viscosity $\eta_m$.  Its torque-induced angular velocity is
then
\begin{equation}
  \vec{\Omega}(\vec{r})
  =
  \frac{\boldsymbol{\tau}(\vec{r})}{8 \pi \eta_m a_p^3},
\end{equation}
which corresponds to a rotation rate on the order of
$\Omega = \SI{1}{\hertz}$ for the present system.  This estimate
suggests that $\Omega < \Omega_c$ for TPM spheres in water, which is
consistent with the observation that solid TPM spheres do not form
chains.  The rotation rate is likely to be substantially increased by
the deformability of the fluid droplets \cite{nadal2014asymmetric},
particularly if resonances enhance their influence
\cite{dileonardo2007parametric}.  A formulation of this enhancement is
not yet available, however, which means that the observed rapid
rotation of acoustically levitated fluid droplets is an outstanding
challenge.

\section{Polar orientation of co-rotating spheres}
\label{sec:polarAngle}

The polar orientation, $\theta$, of a pair of corotating spheres
evolves in time because of three influences: (1) hydrodynamic
interactions that tend to make the spheres tumble out of plane, (2)
the primary acoustic force that drives them back toward the nodal
plane, and (3) secondary Bjerknes interactions.  From
Sec.~\ref{sec:hydrodynamicforces}, the hydrodynamic contribution to
$\dot{\theta}$ is
\begin{equation}
  \dot{\theta}_h
  =
  2\Omega \, \left(\frac{a_p}{r}\right)^3 \, \cos \phi.
\end{equation}
The primary acoustic force $F_1(\vec r)$ contributes
\begin{align}
  \gamma \dot\theta_1
  & =
    \frac{k \, p_0^2}{4r} (\beta'-\alpha') \sin\theta \sin(2kr\cos\theta) \\
  & \approx
    \frac{1}{4} (k p_0)^2 \, (\beta'-\alpha')
    \, \sin(2 \theta),
\end{align}
and the contribution due to acoustic interactions,
$\vec{F}_2(\vec{r}_1, \vec{r}_2)- \vec{F}_2(\vec{r}_2, \vec{r}_1)$, is
\begin{multline}
  \label{eq:theta_a}
  \dot{\theta}_2 = \frac{p_0^2}{2 c_m^2 \rho_m \eta_m}
  \frac{a_p^5}{r^5} \, \sin(2\theta) \bigg[ f_1^2
  + \frac{f_1(8f_0 + 9f_1)}{36} (kr)^2\\
  +\frac{(2 f_0 + 3 f_1)^2}{108} (kr)^4
  +\frac{2 f_0 + 3 f_1}{18 (ka_p)^3} (kr)^5\\
  -\frac{16 f_0^2 + 24 f_0 f_1 + 15 f_1^2}{864}(kr)^6 \bigg].
\end{multline}
For the stable point analysis in Sec.~\ref{sec:acoustohydrochaining},
the pair of interacting spheres is assumed to be in contact,
$r = 2 a_p$, because of attractive acoustic interactions.  Recognizing
that $k r \approx 2 k a_p \ll 1$, we retain only the leading term of
$\dot{\theta}_2$ when deriving Eq.~\eqref{eq:Omega_c}.

%

\end{document}